\documentclass[conference]{IEEEtran}
\usepackage[utf8]{inputenc}
\usepackage{CJKutf8}
\usepackage{comment}
\usepackage{amsmath}
\usepackage{tabularx}
\usepackage{booktabs}
\usepackage{multirow}
\usepackage{mathtools}
\usepackage{graphicx}
\usepackage{amssymb, mathtools}
\usepackage[hidelinks]{hyperref}
\usepackage{capt-of}
\usepackage[pass]{geometry}
\usepackage{xcolor}
\usepackage{balance}
\usepackage{CJKutf8}
\usepackage{float}
\usepackage{tabularx}
\usepackage{colortbl}
\usepackage{xcolor}

\usepackage{algorithm}
\usepackage{algorithmic}
\usepackage{titlesec}

\AtBeginDocument{
  \setlength{\abovedisplayskip}{6pt}
  \setlength{\belowdisplayskip}{6pt}
  \setlength{\abovedisplayshortskip}{3pt}
  \setlength{\belowdisplayshortskip}{3pt}
  \setlength{\textfloatsep}{6pt}   
\setlength{\floatsep}{6pt}       
\setlength{\intextsep}{6pt}      
}

\input epsf
\usepackage{graphicx}

\begin{document}

%



\title{Cyberspace Search Intentions as Leading Indicators for Proactive Traffic Hotspot Detection
}





\author{
    \IEEEauthorblockN{Hangli Ge\IEEEauthorrefmark{1}, Dizhi Huang\IEEEauthorrefmark{2}, 
    Takeshi Kawasaki\IEEEauthorrefmark{3},
    Noboru Koshizuka\IEEEauthorrefmark{1}}
    \IEEEauthorblockA{
\IEEEauthorrefmark{1}Interfaculty Initiative in Information Studies; }
\IEEEauthorblockA{\IEEEauthorrefmark{2}Graduate School of Interdisciplinary Information Studies, The University of Tokyo\\
\IEEEauthorrefmark{3}Operation Division ITS Promotion Department, East Nippon Expressway Company Limited, Japan} \{hangli.ge,  dizhi.huang, noboru\}@koshizuka-lab.org
\IEEEauthorrefmark{3}t.kawasaki.ab@e-nexco.co.jp
}

\makeatletter
\def\blfootnote{\xdef\@thefnmark{}\@footnotetext}
\makeatother

\maketitle

\blfootnote{This work has been accepted for publication in the proceedings of the IEEE International Conference on Systems, Man, and Cybernetics (IEEE SMC 2026). Copyright may be transferred without notice, after which this version may no longer be accessible.}

\begin{abstract}


This study proposes a cyber-physical data-driven framework for proactive detection of highway traffic hotspots and hot regions. The proposed framework bridges users’ online search records in cyberspace as early indicator. To handle large-scale and irregular search records, we propose an Origin–Destination–Time (ODT) tensor model to represent the spatio-temporal structure of route search data and accelerate computation. Using destination-wise inflow sequences derived from these records, we develop a systematic method to automatically identify anomalous surges that indicate emerging traffic hotspots and further the regions. To validate the framework, we conduct experiments using a one-year real-world dataset covering 2,728 interchange (IC) nodes within a highway network. Furthermore, we integrate and compare search data with actual traffic volumes for evaluation. The results reveal a strong correlation between search intensity and traffic flow, demonstrating that online search behavior serves as a reliable proxy for anticipating traffic dynamics. These findings suggest that route search records in cyberspace can be effectively utilized for proactive traffic monitoring and highlight the potential for early prediction of congestion patterns.

\end{abstract}



\section{Introduction}
Understanding and predicting traffic demand is essential for intelligent transportation systems, as it directly affects congestion control, traffic safety, transportation planning. Existing approaches primarily rely on observed traffic flows or static geographic features, making them inherently reactive and limited in capturing the underlying drivers' mobility demand. Route search behaviors in cyberspace provide a direct and interpretable proxy for travel intention signals. Especially when users specify departure or arrival times, these logs offer strong indicators for predicting future traffic trends around the specified time. By shifting the perspective from observed traffic conditions to pre-trip search behaviors, we moves beyond reactive analysis toward \textbf{behavior-informed proactive anomaly detection}.
We propose \textbf{a data-driven framework that leverages large-scale expressway search records to capture latent travel intentions and infer emerging hotspots and high-impact regions}.


\begin{figure}[t]
\centering 
\includegraphics[height=1.7in, width=1\linewidth]{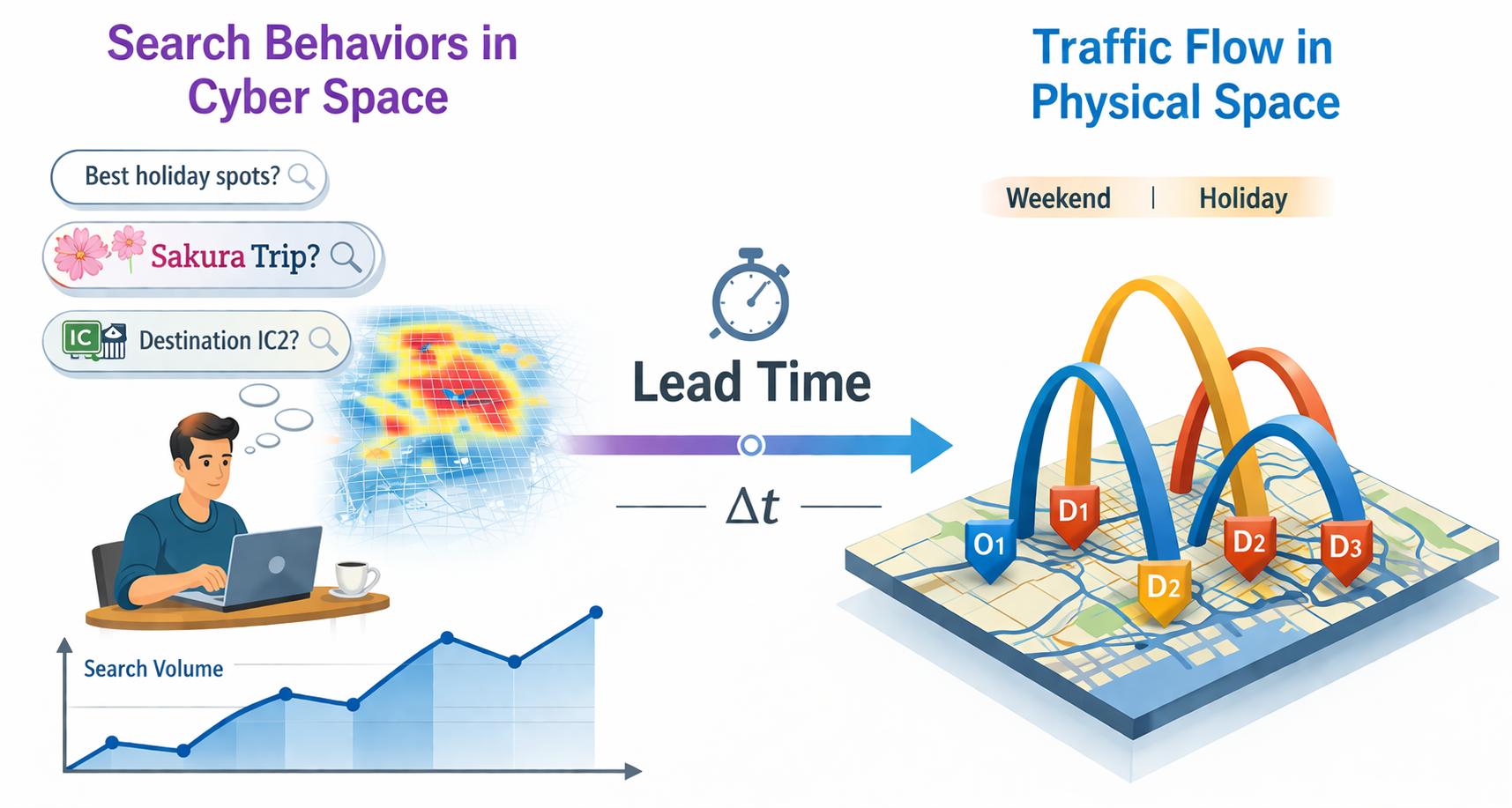} 
\caption{Conceptual illustration of proactive traffic detection using search behavior as a leading indicator, where the lead time ($\Delta t$) captures the gap between intention and observed traffic.} 
\label{img/fig3} 
\end{figure}

Figure~\ref{img/fig3} illustrates the conceptual overview of our proposed framework. It bridges cyberspace behavior and physical mobility, enabling a shift from accumulative search records to traffic  demand modeling. A key insight of our proposal is the existence of a lead time ($\Delta t$) between human intention formation and the realized traffic flow. During this interval, aggregated search signals provide an early indication of emerging demand, which collectively form spatiotemporal patterns of search inflow.


As a proposal, the user route search records in cyberspace are constructed as origin–destination–time (ODT) tensor for spatio-temporal modeling and fast processing~\cite{ge2025origin}. Based on the destination-wise inflow sequences, we compute anomaly scores to identify abnormal search volume patterns and detect spatially coherent hotspots along the geographically ordered destination dimension. These hotspots are further analyzed through temporal and geographical features, including geographical neighborhood to enabling hot region clustering. Experimental analysis on large-scale real-world highway data demonstrates that the proposed approach effectively captures dynamic mobility semantics and reveals interpretable spatiotemporal patterns. The results highlight the importance of integrating human search behaviors with observed traffic flow, validating the effectiveness of search records in cyberspace for inferring the traffic trends, which further providing a foundation for proactive and adaptive traffic management. This study makes the following contributions:

\begin{itemize}
    \item We propose a novel \textbf{search-driven hotspot/hot-region detection} framework that leverages user search behavior as human intention signals.
    
    \item To handle the large scale of highway search records, we developed a scalable data structure which is ODT tensor. It eases destination-level aggregation, temporal anomaly detection, and spatial hotspot extraction along geographically ordered destinations.
    
    \item We conduct experiments using real-world expressway route search records from \textit{DoRaPuRa} (NEXCO East, Japan), a highway route search service, covering a highway network with 2,728 IC nodes. The evaluations demonstrate the effectiveness of our approach in identifying interpretable patterns such as commuting and tourism demand.
\end{itemize}

\section{Related Work}
\subsection{OD/ODT Representation for Mobility Analysis}
Origin–Destination (OD) data provide a fundamental representation for large-scale mobility analysis. Traditional OD models focus on aggregate flow patterns, using matrix-like or thematic representations to reduce visual complexity while preserving spatial structure  \cite{wood2010visualisation, guo2014origin}.  Extensions such as flow density estimation and spatial generalization further enable the extraction of dominant mobility patterns from large datasets \cite{gu2023classification}. Recent work has advanced toward Origin–Destination–Time (ODT) representations, which incorporate temporal dynamics and allow reconstruction of fine-grained traffic evolution from coarse trip records. However, existing OD/ODT  \cite{ge2025origin} representations primarily capture observed mobility (realized flow), while largely overlooking the behavioral layer (search and intention signals) that precedes actual movement. This limits their ability to capture early-stage dynamics and and human intention-driven modeling of traffic demand. 
\subsection{Proactive Traffic Anomaly Monitoring}
Traffic anomaly detection has evolved from statistical methods (e.g., thresholding, residual analysis) to machine learning approaches (e.g., clustering, HMM, GMM), and more recently to deep learning models (e.g., autoencoders, GNNs) that capture complex spatial–temporal patterns \cite{gao2025real, chandola2009anomaly}. While these methods improve robustness, most remain reactive, identifying anomalies only after they occur. Proactive anomaly detection addresses this limitation by leveraging early signals, predictive modeling and proactive control on traffic\cite{ge2019intellevator}. These findings suggest that search behavior can provide valuable signals for anticipating traffic dynamics before they are observed in physical systems. However, previous methods still rely heavily on realized traffic data, with limited integration of human behavioral signals, making it difficult to distinguish early anomaly precursors from normal variability. To overcome this limitation, proactive traffic anomaly detection aims to anticipate crowdedness or congestion using early warning signals (e.g., flow fluctuations, speed variance) and predictive models. However, challenges remain in integrating driver behavior, distinguishing early anomaly signals from normal variability, and linking detection with route decision-making.

\subsection{Human Intention-driven Modeling}
Emerging research further explores proactive prediction by leveraging external signals, including textual and event-related data, to capture early indicators of mobility changes \cite{liang2024exploring}. These findings suggest that traffic dynamics are fundamentally driven by human intentions. In this context, search query data provide a direct and timely proxy for human travel intention, offering a unique opportunity for proactive traffic monitoring. Compared to conventional traffic observations, search behaviors can serve as leading indicators, enabling earlier detection of potential traffic anomalies and demand surges.

\section{Proposal}
This study introduces a behavior-aware, ODT-based framework for proactive traffic analysis. First, instead of treating destinations as static geographic entities, we model \textbf{destination semantics} as emergent properties of behavior-driven temporal patterns derived from route search dynamics. Second, we explicitly leverages \textit{human intention} by modeling route search behavior (pre-trip, anticipatory signals) with ODT-based search flow. This connection enables the identification of \textbf{early-stage anomaly signals} before they manifest in observed traffic flow. Third, to handle large-scale highway search data, we design a \textbf{ODT tensor} that integrates destination-level aggregation, temporal anomaly detection, and hotspot extraction along a geographically ordered destination dimension. This structure supports both computational efficiency and interpretable spatio-temporal analysis.
\begin{figure}[H] 
  \centering
  \includegraphics[height=1.4in, width=0.98\linewidth]{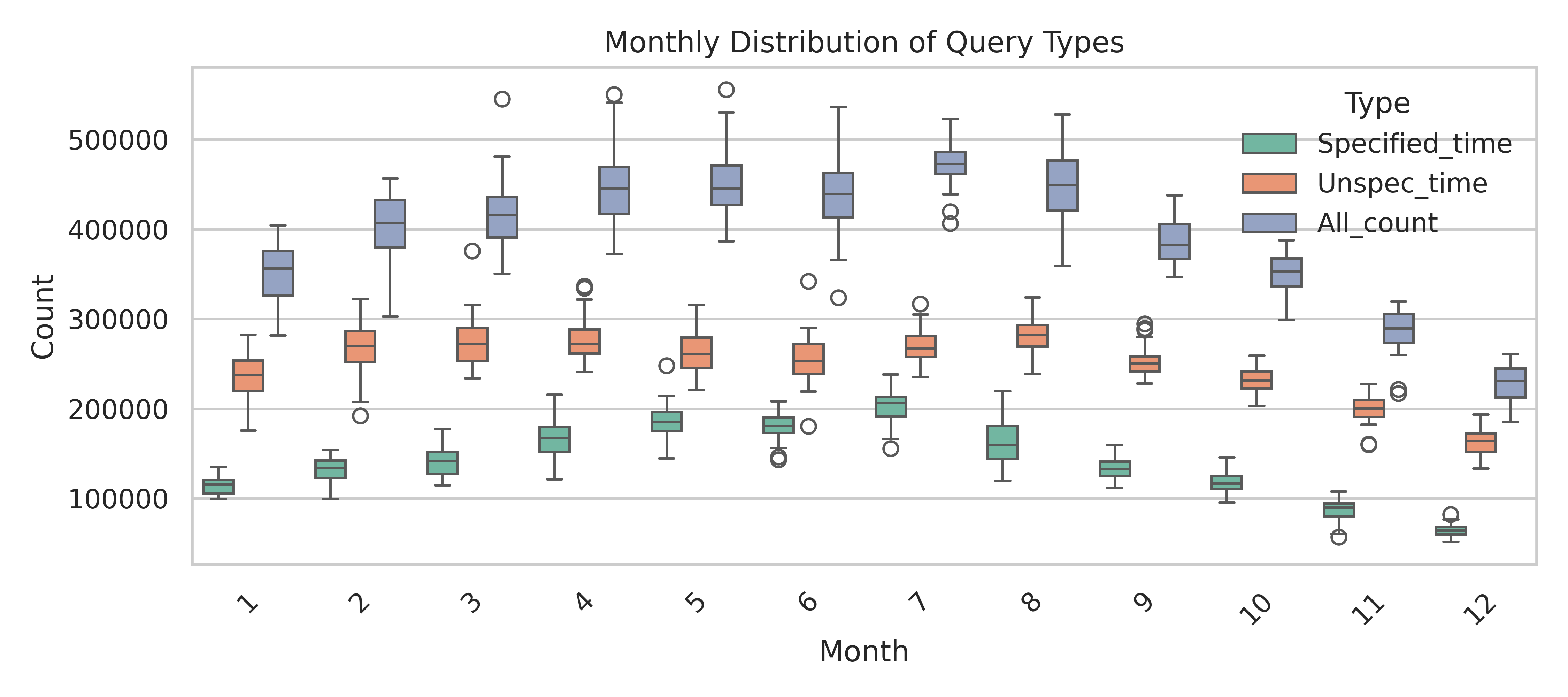}
  \caption{Monthly search volume trends in 2025}
  \label{fig:2025_trends}
\end{figure}

\subsection{Datasets} 
Route search records are collected from the highway service \textit{DoRaPuRa}, where users input origin/destination ICs to query routes, tolls, and travel times. Optional fields include departure/arrival time and vehicle type. Each record contains five attributes: search timestamp, origin IC, destination IC, specified travel time, and travel type. By default, the time field is initialized to the current timestamp. We utilize a dataset from 2025 for evaluation. Figure~\ref{fig:2025_trends} presents the monthly search volume trends, based on over 140 million records, of which more than 51 million contain specified travel times. We utilized the filtered  time specified search records for validation.
\begin{figure}[h] 
  \centering
  \includegraphics[height=1.8in, width=0.9\linewidth]{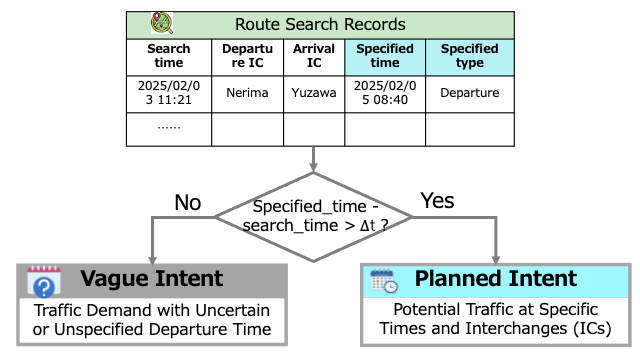}
\caption{Queries are classified into vague or planned intent depending on whether the time is specified (with temporal constraints).}
\label{fig:sepeint}
\end{figure}
As shown in Figure~\ref{fig:sepeint}, queries are separated based on temporal constraints. Since the default time equals the page load time, users who do not modify it typically have a specified time close to the query time. We define a threshold $\Delta t = 30$ minutes: if the specified time exceeds the query time by more than $\Delta t$, the query is classified as \textit{planned intent} (future demand); otherwise, it is labeled as \textit{vague intent}. This enables a clear distinction between planned and uncertain travel behavior.
\subsection{ODT Tensor Construction}
\begin{figure}[H]
\centering 
\includegraphics[height=1.3in, width=0.45\textwidth]{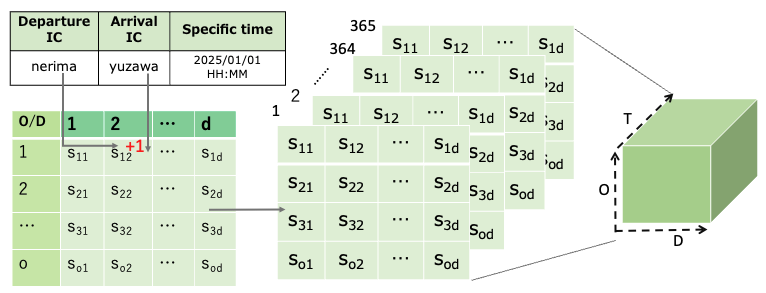} 
\caption{ODT tensor construction process} 
\label{img/odt} 
\end{figure}

Conventional OD maps are typically two-dimensional, capturing flow volumes between origins and destinations. In this study, we extend the traditional OD tensor to tensor by introducing time as a third dimension. Specifically, search records are represented as a 3D tensor 
\(
O \times D \times T \quad \text{(Origin × Destination × Days)},
\)
where the \(O\) and \(D\) axes denote the origin and destination ICs, and \(T\) represents time in days. This formulation enables spatiotemporal analysis of search volume patterns.

Figure~\ref{img/odt} outlines the detailed construction process. First, ICs on the expressway network were assigned sequential numbers based on longitude and latitude. It forms a two dimensional OD matrix of size 2728 x 2728 initialized with zeros. Daily statistical processing was performed, with one two dimensional OD map generated per day. These are stacked over the year, creating a 3D OD map with dimensions  \( O \times D \times T \quad \text{(Origin × Destination × Days)} \). Each search record is processed to calculate the number of days elapsed from January 1st, based on the specified time. Using the origin IC (\(i\)) and destination IC (\(j\)), the corresponding element \(S_{i,j}\) in the OD matrix for the calculated search volume is incremented by one. Repeating this process for all records completes the OD tensor.



\subsection{Hotspots Extraction Along the Destination Dimension}

Given the ODT tensor $\mathcal{X} \in \mathbb{R}^{|O| \times |D| \times |T|}$, we aim to identify geographically coherent anomalous hotspots along the destination dimension.

\textbf{Destination-Level Aggregation:} We first aggregate the ODT tensor along the origin dimension to obtain a destination-level inflow time series:
\begin{equation}
x_d(t) = \sum_{o \in O} \mathcal{X}_{o,d,t}
\end{equation}

\begin{figure*}[t]
\centering 
\includegraphics[height=1.8in, width=1\linewidth]{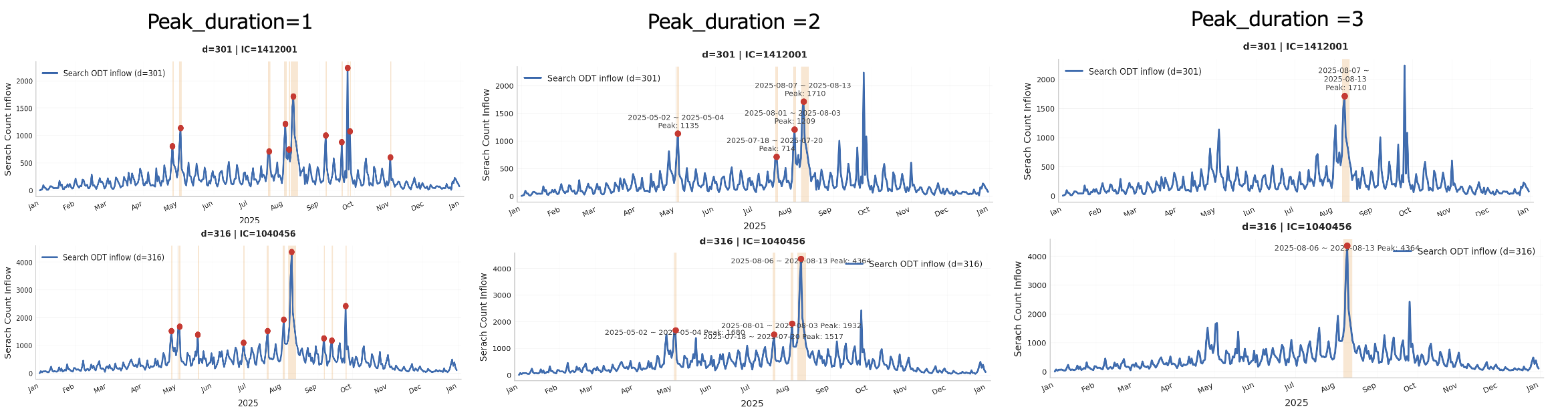} 
\caption{Several examples of inflow at selected destinations ($d = 301$, $d = 336$, and $d = 316$) for Peak\_duration = 1, 2, 3 are presented. To reduce visual clutter, text annotations for Peak\_duration = 1 are omitted.}
\label{fig:example-line}

\end{figure*}
\begin{figure*}[t]
\centering 
\includegraphics[height=1.9in, width=1\linewidth]{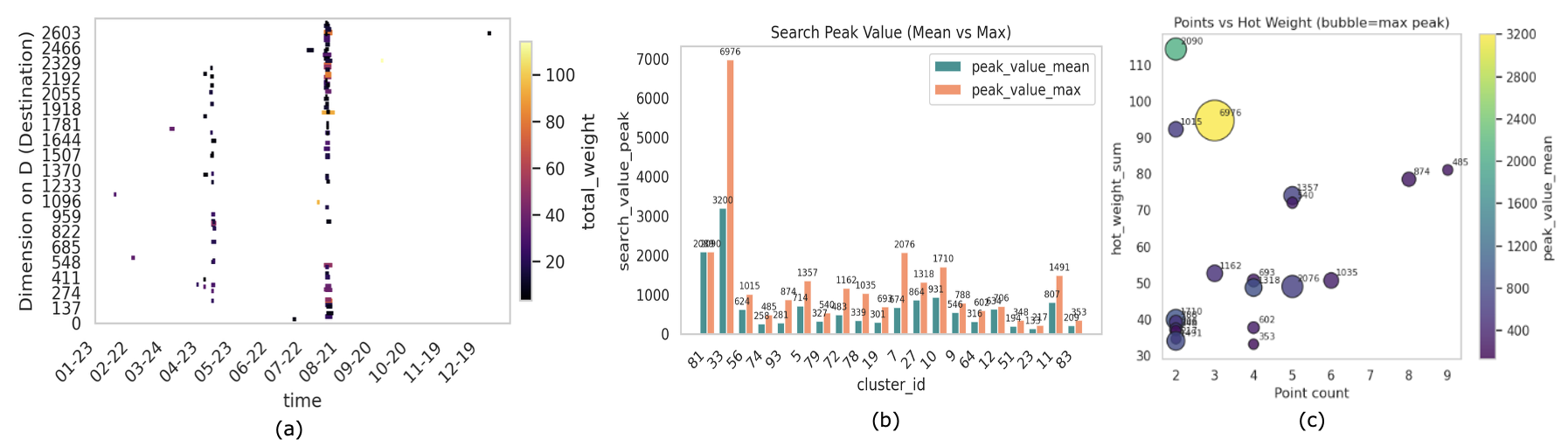} 
\caption{
Region-level clustering analysis. 
(a) Temporal distribution across destination dimension $D$, with color indicating total weight. 
(b) Mean (green) vs.\ maximum (orange) peak values per cluster, highlighting variability and extreme events. 
(c) Relationship between point count and hot weight, where bubble size represents maximum peak value.
}
\label{fig:3pics}
\end{figure*}

\textbf{Hotspot Candidate Selection:} To detect abnormal inflow patterns, a destination-wise anomaly score $z_d(t)$ is computed by comparing observed inflow with temporally conditioned baseline status:
\vspace{-6pt}
\begin{equation}
z_d(t)=\frac{x_d(t)-\hat{x}_d(t)}{\sigma_d}
\end{equation}
where $\hat{x}_d(t)$ represents the expected inflow under normal conditions, and $\sigma_d$ is the standard deviation.
For each destination \(d\), an adaptive anomaly threshold is defined using a quantile threshold $\tau_d$. 
\vspace{-6pt}
\begin{equation}
\tau_d = Q_q\!\left(\{z_d(t)\}_t\right),
\end{equation}
where \(Q_q(\cdot)\) is the \(q\)-quantile (e.g., \(q=0.9, q=0.95\), etc.).  The anomalous time set $\mathcal{S}_d$ for destination \(d\) is:
\begin{equation}
\mathcal{S}_d = \{ t \mid z_d(t) > \tau_d \}, 
\label{eq4}
\end{equation}
Consecutive time indices in \(\mathcal{S}_d\) are grouped into anomalous segments:
\begin{equation}
S_k(d)=\{t_s,\dots,t_e\},\;
\ell_i=t_e-t_s\ge L_{\min},\;
w_i=z_i\ell_i
\label{eq5}
\end{equation}

where \(L_{\min}\) is the predefined minimum segment duration for filtering. As shown in Equation \ref{eq5}, each hotspot is assigned a weight defined as the product of $z_i$ and $\ell_i$, reflecting both its intensity and temporal persistence.
\subsection{Spatiotemporal Hotspot Region Clustering}
\label{sec:st_hotspot}
We aim to identify coherent hotspot regions in the joint space of destination and time based on high-intensity anomaly signals. Each record $d_i$ is characterized by a destination location $d(\phi_i,\lambda_i)$ (latitude and longitude). To efficiently identify spatiotemporal clusters, we adopt a local neighborhood-based approach. 
Since all destinations are pre-ordered according to their geographic coordinates, nearby elements in the sorted sequence tend to be spatially adjacent. 
After sorting records by $d(\phi_i,\lambda_i)$, each point $i$ only examines its $K$ nearest neighbors in the sorted order. This ordering enables an efficient approximation of spatial proximity without explicit distance computation. An edge is established between two points $i$ and $j$ if: $|t_i - t_j| \le \Delta_t \quad \text{and} \quad |d_i - d_j| \le \Delta_d.$ It provides a scalable approximation to density-based clustering by restricting connectivity to local neighborhoods.


\textbf{Connected Component Extraction:}
We model the clustering process as finding hot region in a graph, where nodes represent records and edges indicate pairwise connectivity. 
To efficiently maintain and merge components, we adopt a Union-Find structure with path compression and union-by-rank heuristics. Specifically, for each valid pair $(i,j)$, we perform $\texttt{union}(i,j)$, and cluster assignments are obtained by grouping elements with the same root $\texttt{find}(i)$. 
This yields an amortized time complexity of $\mathcal{O}(\alpha(n))$ per operation, where $\alpha(\cdot)$ is the inverse Ackermann function. This clustering solution reduces computational complexity while preserving local structure.
Each cluster $r$ corresponds to a set of spatiotemporal segments. We retain only clusters with cardinality satisfying $|R_c| \ge N_{\text{limit}}$ (with $N_{\text{limit}} \ge 2$), where
$R_c \subseteq S(d)$.
We then aggregate member records to compute summary statistics, including the number of points $N_r$, the set of destinations $S_d$, mean and maximum anomaly scores ($z_{\text{mean}}$ and $z_{\text{max}}$), total weight $\sum_{i: s(i)\in r} w_i$, geometric center $(\phi,\lambda)$, and   $peak\_duration$ $\ell_r$, defined as
\vspace{-2pt}
\begin{equation}
\ell_r =
\max_{i: s(i)\in r} t_i^{\mathrm{end}} - \min_{j: s(j)\in r} t_j^{\mathrm{start}}.
\end{equation}

\section{Evaluation}
Experiments were performed on an Ubuntu~22.04.5 workstation with an Intel Xeon w5-3435X (16C/32T), 256~GiB  RAM. Code used Python~\texttt{3.10.12} runs used the CPU backend.
\subsection{Hotspot Detection Result }
\begin{figure}[t]
\centering 
\includegraphics[height=2.1in, width=0.8\linewidth]{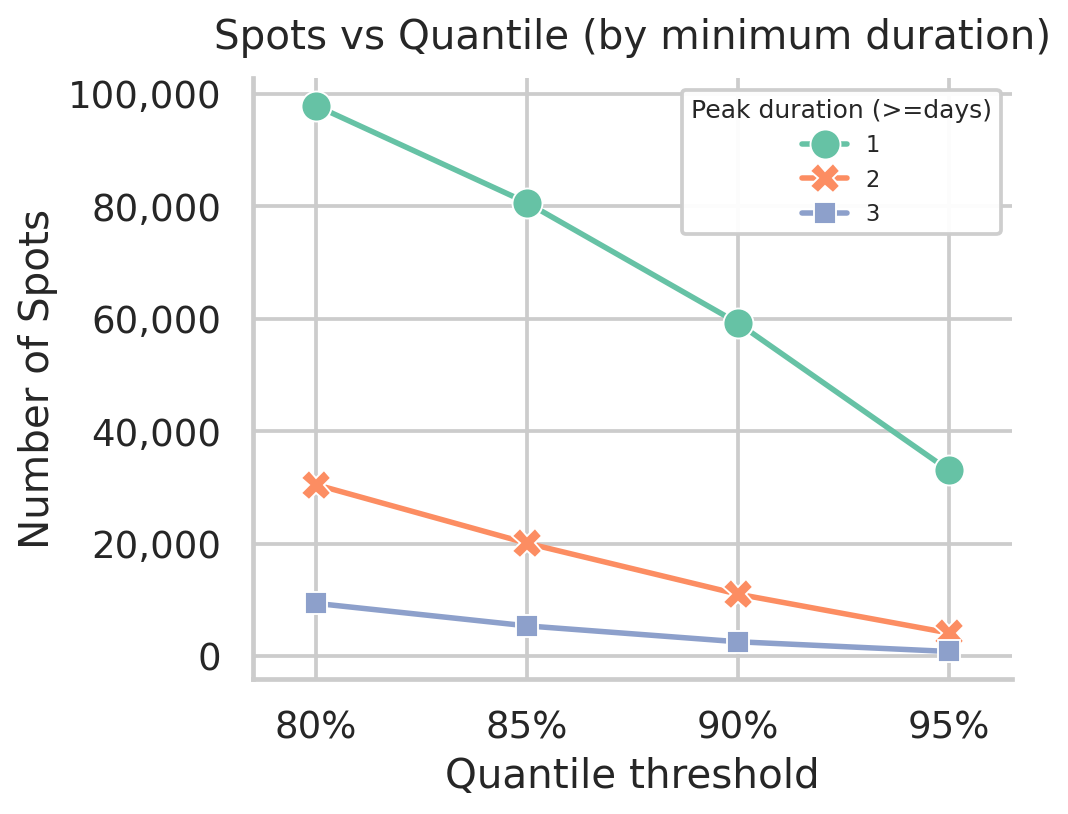} 
\caption{Number of detected hotspot locations under varying quantile thresholds and minimum duration constraints.}
\label{fig:quantile_spots}
\end{figure}
\vspace{-1pt}
\textbf{Overall Statistics:} Figure~\ref{fig:quantile_spots} illustrates the number of detected hotspot locations under different quantile thresholds and minimum duration constraints. As the quantile threshold increases from $80\%$ to $95\%$, the number of detected spots decreases monotonically across all settings. This indicates that higher thresholds effectively filter out moderate anomalies and retain only the most significant events. Furthermore, imposing a stricter minimum duration substantially reduces the number of detected hotspots. For example, requiring persistence of at least 3 days leads to a significant drop compared to 1-day events, suggesting that many anomalies are short-lived. Overall, it demonstrate a clear trade-off between sensitivity and robustness: lower thresholds and shorter durations capture a larger set of transient anomalies, while higher thresholds and longer durations focus on fewer but more persistent and reliable hotspot patterns. 

\textbf{Representative Examples:} Figure~\ref{fig:example-line} illustrates the effect of different peak duration thresholds ($\text{Peak\_duration} = 1, 2, 3 $, left to right) on detecting abnormal search inflow events. The time series exhibits clear temporal patterns, with prominent spikes corresponding to major seasonal periods such as golden week (late April to early May), summer holidays (July–August), and silver week (September). These patterns vary across destinations, reflecting location-specific demand dynamics. The detection results show that shorter durations capture more transient peaks, while longer durations focus on sustained events. As the duration increases, detected peaks become fewer and more concentrated around major seasonal periods, and the shaded intervals become more coherent. However, this also leads to missing smaller but potentially relevant peaks.

\subsection{Region-level Clustering Summarization} 
Figure~\ref{fig:3pics} presents a comprehensive analysis of detected segments from temporal, statistical, and structural perspectives. As shown in Figure~\ref{fig:3pics}(a), each segment is projected onto the destination (y-axis) dimension over time (x-axis). The results reveal sparse yet highly bursty patterns, with segments concentrated in specific time windows. The color intensity indicates total weight, highlighting that high-intensity events are temporally localized and often involve multiple destinations simultaneously. It suggests strong event-driven behaviors and cross-destination correlations.

\textbf{Statistical characteristics:} Figure~\ref{fig:3pics} presents the temporal, statistical, and structural characteristics of detected segments. 
Temporally (a), segments are sparse but highly bursty, with high-intensity events concentrated in specific periods and often spanning multiple neighborhood destinations. Statistically (b), the gap between mean and maximum peak values indicates the presence of rare but high-impact events.  Structurally (c), clusters with few points exhibit high intensity, revealing a decoupling between event frequency and impact. Overall, the results highlight bursty dynamics, inter-cluster variability, and intention-driven hotspot evolution.
\begin{figure}[t]
\centering 
\includegraphics[height=1.9in, width=0.8\linewidth]{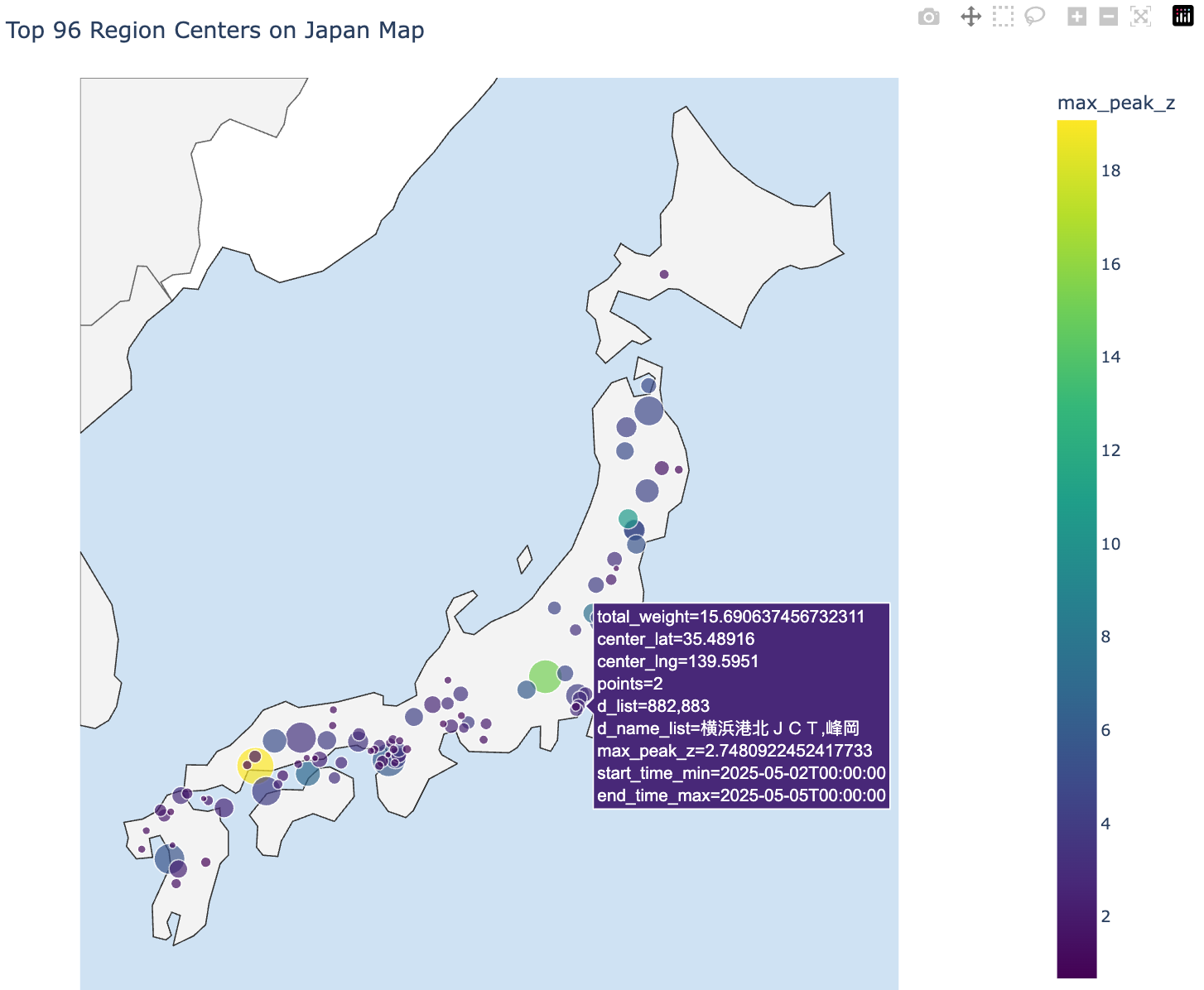} 
\caption{Interactive visualization of top region centers on a map of Japan. }
\label{fig:map_plot}
\end{figure}

\textbf{Geographical Distribution:} We further develop an interactive map visualization (Figure~\ref{fig:map_plot}) to explore region clusters geographically across Japan. 
Circle size represents total weight and color encodes peak intensity ($z$-score). 
High-intensity clusters concentrate in urban and tourist regions, while rural areas remain sparse. 
The interactive interface enables detailed inspection and reveals strong spatial heterogeneity in regional demand patterns.

\subsection{Validation: Correlation Analysis with Traffic Volume}
To validate the effectiveness of the proactive search records driven framework, we examine the relationship between search volume and traffic flow. Specifically, we quantify the association between time specified searches for destination interchanges (ICs) and inflow traffic volume (number of vehicles) on the corresponding ICs. The Pearson correlation coefficient was utilized, with the equation~\ref{eq7} defined as follows.
\begin{equation}
\begin{split}
\rho_d &= \frac{\sum_{t} (q-\bar{q}_d)(v-\bar{v}_d)}
{\sqrt{\sum_{t} (q-\bar{q}_d)^2}
 \sqrt{\sum_{t} (v-\bar{v}_d)^2}}
\end{split}
\label{eq7}
\end{equation}
\vspace{-2pt}
Let $q(t, d)$ denote the search inflow count for  destination IC $d$ at time $t$, and $v(t, d)$ denote the observed inflow traffic volume at destination $d$ at time $t$ over time, $\bar{q}_d$ and $\bar{v}_d$ denote the mean values of $q(t,d)$ and $v(t,d)$ respectively. The correlation coefficient ranges from $-1$  to $1$. We focus on two major highways of Japan, the Kan-Etsu Expressway (E17) and the Tohoku Expressway (E4), covering 46 interchanges (ICs). Traffic counter data, sampled at an hourly resolution, is utilized for evaluation. 
\vspace{-4pt}
\begin{table}[htbp]
\centering
\caption{Top-5 ICs with the highest Pearson correlation}
\label{tab:quarter_reorg}
\resizebox{\linewidth}{!}{
\begin{tabular}{c|cc|cc|cc|cc|cc}
\hline
\multirow{2}{*}{Rank} 
& \multicolumn{2}{c|}{Full Year} 
& \multicolumn{2}{c|}{Q1} 
& \multicolumn{2}{c|}{Q2} 
& \multicolumn{2}{c|}{Q3} 
& \multicolumn{2}{c}{Q4} \\
\cline{2-11}
& IC & Corr. & IC & Corr. & IC & Corr. & IC & Corr. & IC & Corr. \\
\hline
1 & \cellcolor{blue!10}\textbf{1800106} & \cellcolor{blue!10}\textbf{0.789}
  & \cellcolor{blue!10}\textbf{1800066} & \cellcolor{blue!10}\textbf{0.747}
  & \cellcolor{blue!10}\textbf{1800106} & \cellcolor{blue!10}\textbf{0.888}
  & \cellcolor{blue!10}\textbf{1800111} & \cellcolor{blue!10}\textbf{0.918}
  & \cellcolor{blue!10}\textbf{1800111} & \cellcolor{blue!10}\textbf{0.873} \\

2 & 1800086 & 0.781
  & 1800056 & 0.745
  & 1800086 & 0.873
  & 1800106 & 0.912
  & 1800061 & 0.847 \\

3 & 1040076 & 0.770
  & 1800086 & 0.745
  & 1040076 & 0.870
  & 1040051 & 0.866
  & 1800106 & 0.841 \\

4 & 1800071 & 0.758
  & 1800071 & 0.738
  & 1800066 & 0.831
  & 1800056 & 0.848
  & 1800066 & 0.840 \\

5 & 1800066 & 0.755
  & 1800061 & 0.728
  & 1800071 & 0.823
  & 1040086 & 0.836
  & 1800051 & 0.837 \\
\hline
\end{tabular}
}
\label{tab:quarter_vertical}
\end{table}

\begin{figure*}[t]
\centering 
\includegraphics[height=2.9in, width=0.9\textwidth]{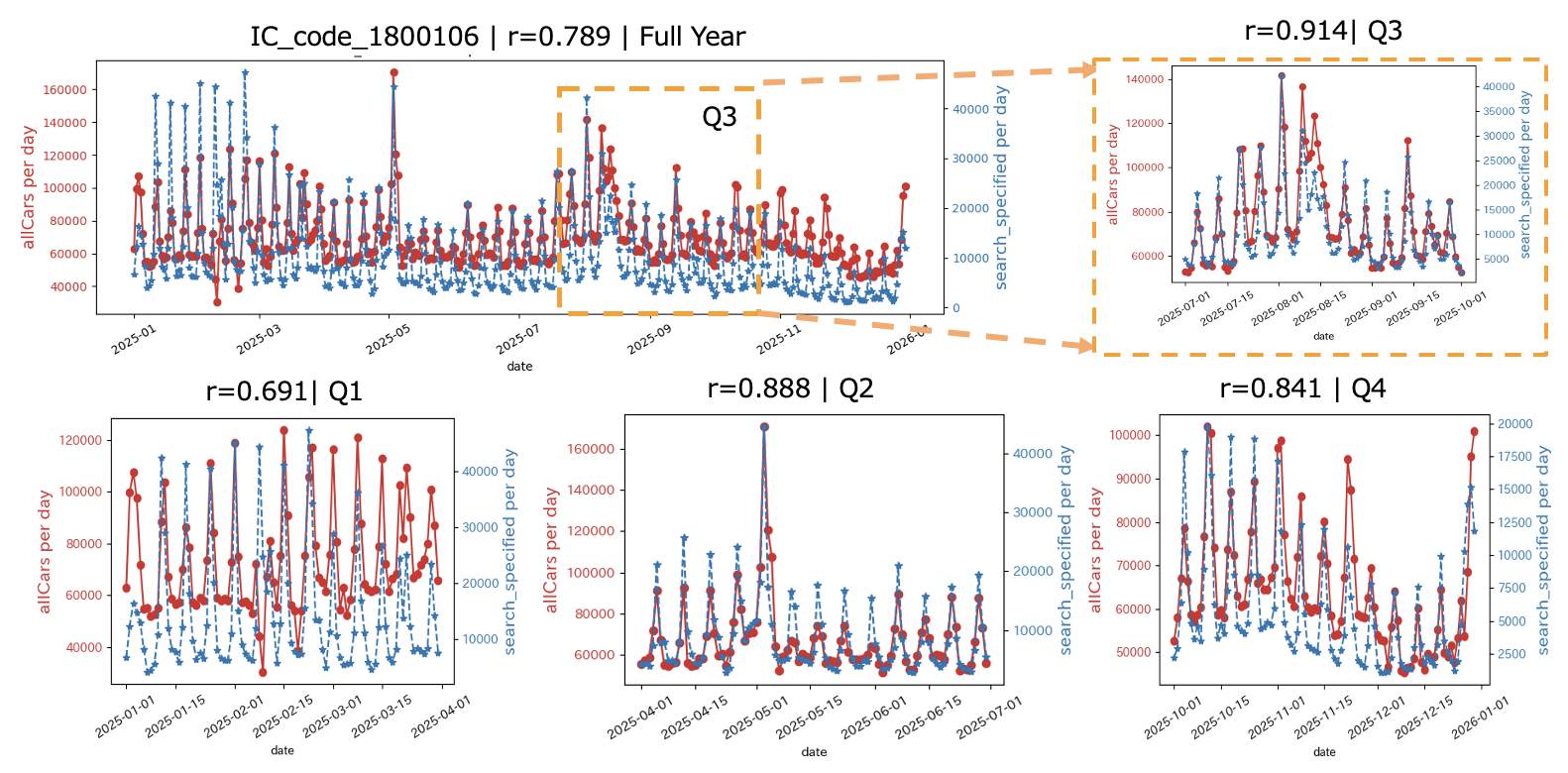} 
\caption{Example of correlation between search volume and traffic flow at IC 1800106.}
\label{fig:corre}
\end{figure*}
\vspace{-4pt}
Table~\ref{tab:quarter_vertical} reports the Top-5 ICs (listed by the code) with the highest Pearson correlation between traffic inflow and search demand across the full year and each quarter. Overall, correlations remain consistently high (mostly $>0.75$), indicating a strong alignment between search behavior and actual traffic. IC 1800106 shows the most stable correlation, ranking first for the full year (0.789) and remaining among the top in Q2--Q4, consistent with its strong temporal patterns observed earlier, achieves the highest correlations (up to 0.918). IC 1800111 dominates in Q3 and Q4, highlighting its importance as a key hotspot in the second half of the year. In contrast, Q1 exhibits relatively lower correlations ($0.73$--$0.75$), implying higher variability and weaker predictability, likely due to seasonal effects of winter. Figure~\ref{fig:corre} further compares daily traffic volume (allCars) and search demand for specific IC codes across the full year and by quarter. Overall, \textbf{the two signals are positively correlated}, meaning search demand generally reflects actual traffic.
\vspace{-1pt}
\section{Conclusion}
\vspace{-1pt}
In this study, we model route search behaviors in cyberspace as leading indicators of future traffic demand in physical space. We propose a novel search behavioral framework for traffic hotspot and region detection from human intention signals. By constructing a origin–destination–time (ODT) tensor from large-scale route search logs, our approach captures the dynamics of search behavior as proactive signals for monitoring future traffic flow. Experimental results on real-world highway data validate the effectiveness of modeling the between intention formation and realized mobility, demonstrating the capability of detecting emerging demand before it appears in physical traffic.

Beyond traffic analysis, our findings highlight the broader value of leveraging intention signals for applications such as congestion prediction and management \cite{kosugi2022traffic,matsunaga2023improving,ge2024frtp, lin2025diffusion}, tourism analysis ~\cite{ge2025origin} and disaster response. By bridging cyberspace behavior and real-world mobility, this work contributes to the development of proactive, adaptive, and human-centric intelligent transportation systems. Future work will focus on deeper integration of multi-source data and enhancing real-time decision support, further advancing the role of large-scale cyberspace data in traffic modeling and management.

\section{Acknowledgement}
This study was conducted under a joint research project between the University of Tokyo and the East Nippon Expressway Co., Ltd. (NEXCO East). This study was also supported by JSPS KAKENHI Grant Number JP25K21205. The data used in this study, such as historical online search logs, and road structure information, were provided by NEXCO East. We gratefully acknowledge the kind support provided by the NEXCO East.

\balance
\bibliography{IEEEabrv,aaai22}

\bibliographystyle{IEEEtran}

\end{document}